\documentclass[12pt]{iopart}

\usepackage{iopams}
\usepackage{graphicx}
\usepackage{cite}
\usepackage[active]{srcltx}

\begin{document}

\title[]{Modeling of continuous absorption of electromagnetic radiation in dense partially ionized plasmas}

\author{A A Mihajlov$^1$, N M Sakan$^1$, V A Sre{\' c}kovi{\' c}$^1$ and Y Vitel$^2$}

\address{$^1$ Institute of Physics, Belgrade University, Pregrevica 118,
Zemun, 11080 Belgrade, Serbia}

\address{$^2$ UPMC Univ Paris 6, Laboratoire des Plasmas Denses, 3 rue
Galilee, 94200 Ivry sur Seine, France.}
\ead{mihajlov@ipb.ac.rs}

\begin{abstract}
A new modeling way of describing the continuous absorption of 
electromagnetic (EM) radiation in dense partially ionized hydrogen
plasma is tested in this work. It is shown that the obtained results
give a possibility of calculating spectral absorption coefficients
which characterize the relevant absorption processes in partially
ionized hydrogen plasmas with electron densities $N_{e} \sim
10^{19}cm^{-3}$ and temperatures $T \approx 2\cdot 10^{4}K$. A key
parameter of the used procedure is determined empirically. The
calculation method is applied to wavelength region $300nm < \lambda
< 500nm$. The presented results can be of interest for dense
laboratory plasmas as well as for partially ionized layers of
different stellar atmospheres.
\end{abstract}


\section{Introduction}
\label{sec:in}

By now, direct methods of determination of various plasma
characteristics, based on quantum or classical statistical
mechanics, have been developed only for practically fully ionized
plasmas \cite{kob95,ada94,ich87,ebe76, kra86, rin88, for89, mih04}.
In the case of dense partially ionized plasmas, where the density of
neutral particles (atoms) is close to the density of positively
charged particles (ions), such rigorous methods do not exist at
present. Recently, in \cite{sre10}, this problem was discussed in
connection with the transport properties of dense partially ionized
plasmas. As for their optical characteristics, it is enough to
remind that adequate calculation methods exist only for weakly and
moderately ionized plasmas with electron density $N_{e} \lesssim
10^{17} cm^{-3}$. As it is well known, the influence of the
neighborhood on an exited atom can be neglected in such plasmas, as
for example in the Solar photosphere \cite{mih78, mih07a}, or be
treated as a small perturbation and described within the framework
of a perturbation theory \cite{kob95, dam10, oma07, sea94, dim96,
tka97, rei03, min06, SCCS98}.

In this paper we will consider the continuous absorption of EM
(electromagnetic) radiation in dense partially ionized plasma, with
electron density $N_{e} \sim 10^{19} cm^{-3}$, temperature $T
\approx 20 000K$ and atom density $N_{a}\approx N_{e}$. Plasmas with
similar parameters are of interest from both the laboratory
\cite{gav01, dya06} and the astrophysical aspect. Here we keep in
mind the plasma of the inner layers of the solar atmosphere, as well
as of partially ionized layers of other stellar atmospheres, for
example the atmospheres of DA and DB white dwarfs with effective
temperatures between $10 000 K$ and $20 000 K$ (see \cite{koe97,
kie04,fin97}).

Due to the exceptional simplicity of the hydrogen atom, this
research is starting with the hydrogen case. Under the mentioned
conditions the continuous absorption of EM radiation in hydrogen
plasmas are determined by the following radiative processes:
\begin{equation}
    \label{eq:ph}
    \varepsilon _{\lambda} + H^{*}(n,l) \to H^{+} + e_{k'},
\end{equation}
\begin{equation}
    \label{eq:inbs}
    \varepsilon _{\lambda} + e_{k} + H^{+} \to e_{k'} + H^{+},
\end{equation}
\begin{equation}
\label{eq:eH} \varepsilon_{\lambda} + \left\{
\begin{array}{l}
e_{k} + H(1s),
\\
H^{-}(1s^{2}),
\end{array}
\right. \to e_{k'} + H(1s),
\end{equation}
\begin{equation}
\label{eq:HH} \varepsilon_{\lambda} + \left\{
\begin{array}{l}
H^{+} + H(1s),
\\
H_{2}^{+}(1\Sigma^{+}_{g}),
\end{array}
\right. \to H(1s) + H^{+},
\end{equation}
where $\epsilon_{\lambda}$ is the energy of a photon with wavelength
$\lambda$, $n$ and $l$ - the principal and the orbital quantum number of
hydrogen-atom excited states, $e_{k}$ and $e_{k'}$- free electrons
with energies $E = \hbar^{2}k^{2}/2m$ and $E' = \hbar^{2}k'^{2}/2m$
respectively, $m$ the electron mass and $\hbar$ - Plank's constant.

In the considered region of $N_{e}$ and $T$ the processes of
electron-$H^{+}$ and electron-$H(1s)$ inverse bremsstrahlung,
photo-detachment of the negative ion $H^{-}(1s^{2})$,
$H(1s)$-$H^{+}$ absorption charge exchange and photo-dissociation of
molecular ion  $H_{2}^{+}(1\Sigma^{+}_{g})$, which are described by
Eqs.~(\ref{eq:inbs}), (\ref{eq:eH}) and (\ref{eq:HH}), can be
treated in the same way as in previous papers \cite{mih93,mih94}.
Therefore, photo-ionization processes (\ref{eq:ph}) will be in the
focus of attention in the next section, and here let us note only
that in this field methods obtained by extrapolation (with minor
modifications) of the methods developed for weakly and moderately
non-ideal plasmas \cite{dya98,gav01,vit04} have been used for such
processes up until now. Moreover, for determination of absorption
coefficients characterizing their influence in the region of $N_{e}
\gtrsim 10^{19} cm^{-3}$ methods based on Cramer's approximation
\cite{dya06,mof08} have been used so far. So, developing a new
modeling way of describing the continual absorption in dense
partially ionized plasmas, which is the main aim of this paper, is
still an actual task.

We take as the landmarks hydrogen plasmas which were experimentally
studied in \cite{vit04}: with $N_e = 6.5\cdot10^{18}cm^{-3}$ and
$1.5\cdot10^{19}cm^{-3}$, and $T =1.8\cdot10^{4} K$ and
$2.3\cdot10^{4}K$ respectively. The presented modeling way is tested
within the optical range of photon wavelengths: $300 nm \le \lambda
\le 500 nm$. A key parameter of the used numerical procedure is
determined empirically. It is shown that this procedure already
allows for determination of the spectral absorption coefficients
characterizing all the relevant absorption processes in dense
partially ionized hydrogen plasmas, at least in the regions $5\cdot
10^{18} cm^{-3} \lesssim N_e \lesssim 5 \cdot 10^{19}cm^{-3}$ and
$1.6 \cdot 10^{4} K \lesssim T \lesssim 2.5\cdot 10^{4} K$.

The material of this paper is distributed over Sections 2 and 3. In
Sec. 2 the following is presented: the approximation of the cut-off
Coulomb potential, together with the reasons for applying just this
approximation in the considered regions of $N_{e}$ and $T$; the way
of obtaining all the partial spectral absorption coefficients, as
well as their corresponding final expressions. In Sec. 3 the results
of the calculations of the partial and the total absorption
coefficients are presented and discussed.

\section{Theory: the spectral absorption coefficients}
\label{sec:th}
\subsection{The approximation of the cut-off Coulomb potential}

Let us note that numerous papers dedicated to development of rather
rigorous methods of ground-state and exited-atom photo-ionization
processes, were published over the past few decades (see for example
\cite{nut69, res75, mse82, faz84, man85, dya98, fan99, sah09}).
Although some of these methods were tested on the examples of
isolated H-like and He atom systems, the main results of those
papers are intended for more complex atoms and for atoms in extreme
EM fields. Therefore we will consider here that an isolated hydrogen
atom $H^{*}(n,l)$ should be described in the same way as in
\cite{bet57}. The appearance of these papers has certainly
encouraged intensive investigation of the atom ionization processes
in plasmas (see for example \cite{wei98, zha04, sah06, shu08, qi09,
sah10, lin10, lin10b}); however, the regions of temperature and
electron density which are considered here ($N_{e} \sim 10^{19}
cm^{-3}$ and $T \approx 20 000K$) have not been examined yet.
Namely, the results obtained in those papers refer to the regions of
extremely large electron density and temperature, where the
non-relativistic way of describing plasma becomes invalid. Besides,
the models of ion screening which are used in the mentioned papers
are not generally appropriate (as it is being shown below) to the
considered regions of plasma parameters. That is why we examine a
new modeling way here, as was already mentioned above.

In the considered hydrogen plasmas processes Eq.~(\ref{eq:ph}) will
present the main difficulties in describing the continuous
absorption, since under the stated conditions the energy of
interaction of an excited atom with its neighborhood reaches the
order of the corresponding ionization potential. It means that these
processes have to be considered a result of radiative transitions in
the whole system "electron-ion pair + the neighborhood", namely
\begin{equation}
\label{eq:infinbf} \epsilon_{\lambda}+(H^{+}+e)_{n,l}+S_{rest}\to
(H^{+}+e)_{E}+S^{'}_{rest},
\end{equation}
where $S_{rest}$ and $S^{'}_{rest}$ denote the rest of the
considered plasma. However, as it is well known, many-body processes
can sometimes be simplified by a transformation to the corresponding
single-particle processes in an adequately chosen model potential.

In this context we will introduce into the consideration the ion Wigner-Seitz
radius $r_{s;i}$, given by
\begin{equation}
\label{eq:si} r_{s;i} = \left (\frac{3}{4\pi N_{i}}\right
)^{\frac{1}{3}}.
\end{equation}
where $N_{i}$ is the $H^{+}$ ion density, and the characteristic
length $r_{c}$ is defined by the relation
\begin{equation}
\label{eq:Upc} U_{p;c}= - \frac{e^{2}}{r_{c}},
\end{equation}
where the quantity $U_{p;c}$ has the meaning of the mean potential
in which a free electron moves in the considered plasma. Then, we
will take into account the fact that, according to the results of
\cite{mih09b}, the characteristic length $r_{c} \cong r_{s;i}$ in
the hydrogen plasmas which are considered here. It means that in
these plasmas the radius of the zone where an ion and an electron
can be treated as an electron-ion pair in Eq.~(\ref{eq:infinbf}) is
just $r_{c}$.

Here, we will follow the previous papers \cite{mih89}, where it was
noted that an adequate ion screening potential has to satisfy two
main requirements:

a) it has to be practically equal to Coulomb potential ($-e^{2}/r$)
in the mentioned zone, i.e. in the region $r < r_{c}$,

b) outside of this zone ($r > r_{c}$) it has to be practically equal
to the mean potential $U_{p;c}$,

\noindent where $r$ is the distance from the considered ion, and $e$
- the modulus of the electron charge. Condition a) is due by the
fact that the averaged potential, created by the rest of the
considered plasma inside the mentioned zone, where the electron of
$(H^{+}+e)$-pair is localized, is equal to zero in the first
approximation, since the rest is neutral as a whole. As for
condition b), it reflects the fact that under such a condition a
realistic, physically acceptable asymptotics is realized for the
electron wave function characterizing $(H^{+}+e)$-pair in the
considered plasma, which cannot be Coulomb-like as in an isolated
electron-ion pair.

\begin{figure}[h!]
\centering
      \includegraphics[width=\columnwidth,
height=0.75\columnwidth]{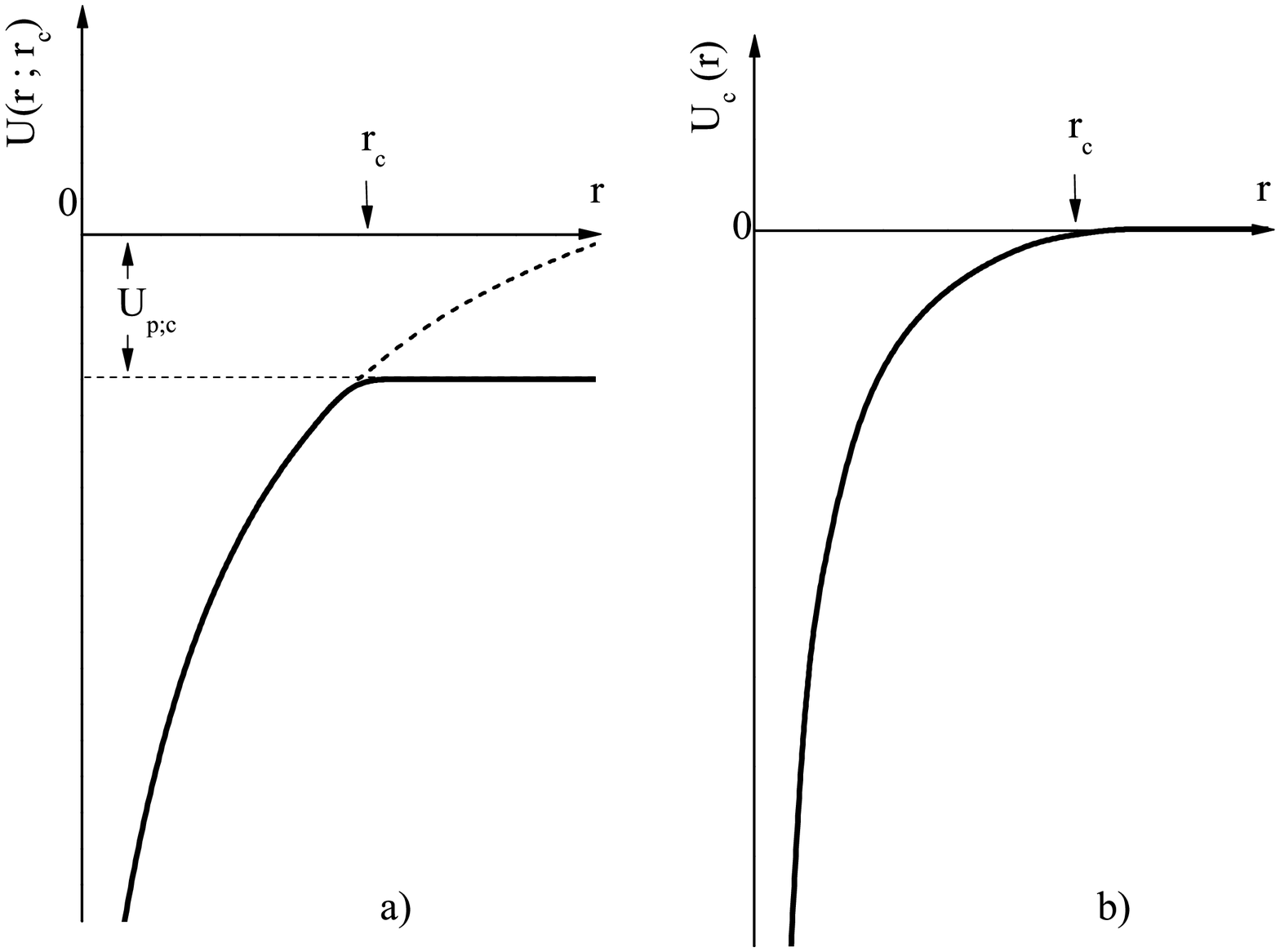} \caption{Potentials $U(r;
r_{c})$ and $U_{c}(r)$, where $r_{c}$ is the cut-off parameter.}
\label{fig::1}
\end{figure}

As an adequate model potential for hydrogen plasma with $N_{e}\sim
10^{19}$ $cm^{-3}$ we choose, as in \cite{mih89}, the screening
cut-off Coulomb potential, which satisfies conditions a) and b), and
can be presented in the form
\begin{equation}
\label{eq:U} U(r;r_{c}) = \left\{
\begin{array}{l}
- \displaystyle \frac{e^2}{r},  \quad  0 < r \le r_{c},
\\
\displaystyle \quad U_{p;c}, \qquad       r_{c} < r < \infty,
\end{array}
\right. ,
\end{equation}
\noindent where the cut-off radius $r_{c}$ is defined by relation
(\ref{eq:Upc}), as it is illustrated by Fig. \ref{fig::1}.a. This
potential was first introduced to the plasma physics in \cite{suc64}
and its properties were investigated later in \cite{kra83,mih86}.

In connection with such a choice of the ion screening potential let
us note the following: the argumentation from \cite{mih89} is taken
into account here, that the often used model of Debye-H{\" u}ckel
(DH) potential is not adequate for description of an electron-ion
pair in dense non-ideal plasma. It is important that we focus here
only on the physical meaning of DH potential, leaving out of the
consideration all of its disadvantages, which have been noted and
discussed in \cite{mih09b, mih08}; namely, as it is well known, DH
potential has the meaning of an average electrostatic potential
created by the observed ion and all the charged particles from the
rest of the plasma. This potential was introduced in \cite{deb23} in
order to describe (in accordance with its meaning) some of the
thermodynamical characteristics of free-particle systems with
Coulomb interaction. From this aspect, its applications have been
justified until now, if the accuracy of the calculations is not very
important \cite{mih09b}.

Let us note also that in order to reduce errors which are generated
by the application of DH potential, instead of DH screening radius
$r_{D}=[kT/(8 \pi N_{e}e^2)]^{1/2}$ for two-component (electron-ion)
plasma, another radius $r_{e}$ is often taken, given by

\begin{equation}
\label{eq:rei} r_{e} = [kT/(4 \pi N_{e} e^2)]^{1/2},
\end{equation}
which in principle characterizes the corresponding single-component system. In
\cite{mih09b} it was shown that replacing $r_{D}$ with $r_{e}$ is
really justified
in weakly non-ideal plasmas, since the adequate
screening radius is approaching just $r_{e}$ when the
non-ideality degree is approaching zero. Here, it is necessary to
remind that by replacing DH screening radius $r_{D}$
with the electron screening radius $r_{e}$ DH potential loses its physical meaning,
becoming one of the numerous DH-like potentials.

However, due to its physical meaning, DH potential could not be
applied in principle to describing electron-ion scattering in dense
plasmas. Namely, in such plasmas the electron in the considered
electron-ion pair is itself a part of the corresponding ion's
"screening cloud" and thus takes part in the creation of DH
potential. This fact is often being forgotten, but neglecting it
yields an absurd situation in the case of such plasmas where DH
screening radius $r_{D}$ is approaching the ion Wigner-Seitz radius
$r_{s;i}$, given by Eq.~\ref{eq:si}. In such a case the complete ion
"screening cloud" consists practically of one electron from the
considered electron-ion pair.

Just this kind of situation is realized in a plasma of the kind
treated in this paper, i.e. with $N_{e}$ of the order of magnitude
of $10^{-19} cm^{-3}$ and $T$ of about 20 000K, where the ion
Wigner-Seitz radius $r_{s;i}$ is close to the ion screening radius
$r_{e}$ given by (\ref{eq:rei}), making an application of DH
potential completely unacceptable. Let us emphasize the fact that
the same is valid for any other potential which has a similar
meaning as DH.

For all these reasons we consider that DH screening potentials,
which were used in the above mentioned papers \cite{wei98, zha04,
sah06, shu08, qi09, sah10, lin10, lin10b} cannot be accepted for an
application in any region of plasma parameters. As for DH-like
potentials, one of them could be accepted in principle for an
application in some regions of $N_{e}$ and $T$, but only under the
following condition:

-the screening radius in the expression for the used potential
guarantees that the values of the basic parameters of the considered
problem (number and energies of the bound states, elastic scattering
amplitude for the angle equal to zero as a function of the free
electron energy, etc.) are very close to the corresponding
experimental data.

This condition could probably be satisfied on the condition of
existence of reliable experimental data which could be used for
determination of the effective screening radius by means of the
corresponding fitting method, and some rigorous procedure being
included for describing the bound and free states of the electron in
the considered DH-like potential. However, for the considered
plasmas there are no such reliable experimental data in the
literature.

Because of all the things mentioned we consider that the choice of
the potential $U(r;r_{c})$, which is given by Eg.~(\ref{eq:U}), is
the best solution in the case of the considered dense partially
ionized plasmas. Besides, an additional advantage of this model
potential is possibility of obtaining all the needed final results
in a compact analytical form.

\begin{figure}[h!]
\centering
      \includegraphics[width=\columnwidth,
height=0.75\columnwidth]{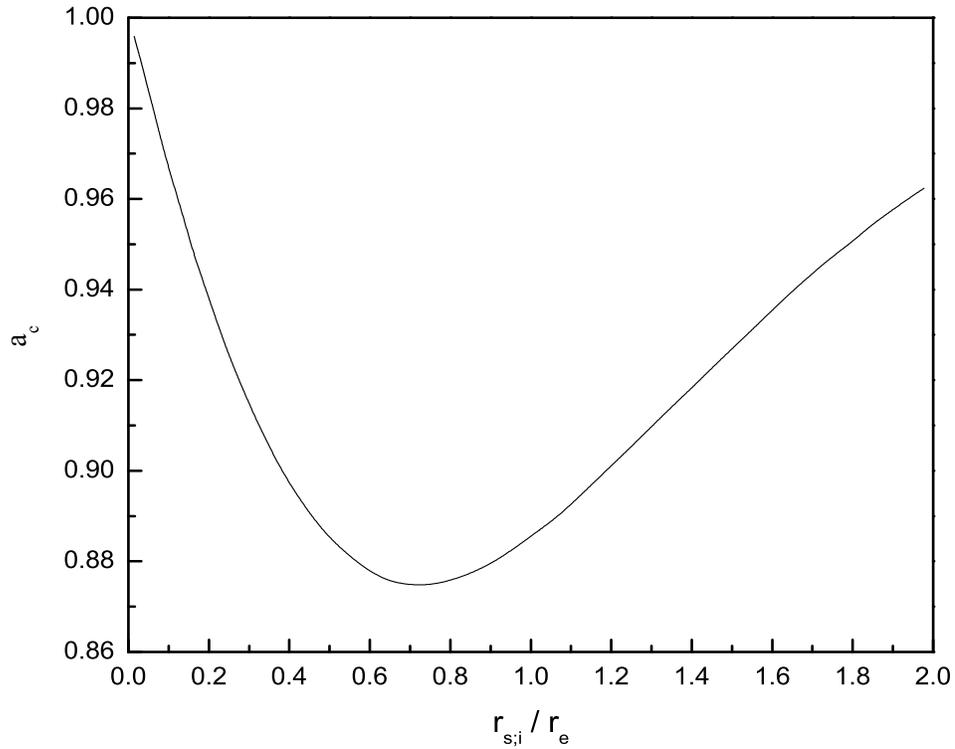}
\caption{The parameter $a_{c}\equiv r_{c}/r_{e}$ as the function of
the ratio $r_{s;i}/r_{e}$, where $r_{s;i}$ and $r_{e}$ are given by
Eqs.~(\ref{eq:si}) and (\ref{eq:rei}).} \label{fig::2}
\end{figure}

As in \cite{mih89}, we will take the value $U_{p;c}$ as the zero of
energy. After that, the potential Eq.~(\ref{eq:U}) is transformed to
the form
\begin{equation}
\label{eq:U0} U_{c}(r) = \left\{
\begin{array}{c c}
- \displaystyle \frac{e^2}{r} + \displaystyle \frac{e^2}{r_c},
\qquad            0 < r \le r_{c},
\\
\displaystyle 0 ,              \qquad      r_{c} < r < \infty,
\end{array}
\right.
\end{equation}
which is used in further considerations, and is also illustrated by
Fig. \ref{fig::1}.b. It is important that, through characteristic
lengths $r_{s;i}$ and $r_{e}$ defined by Eqs.~(\ref{eq:si}) and
(\ref{eq:rei}), the cut-off radius $r_{c}$ is determined here as a
given function of $N_{e}$ and $T$. Namely, taking that
\begin{equation}
\label{eq:pci} r_{c} = a_{c}\cdot r_{e},
\end{equation}
we can directly determine quantity $a_{c}$ as a function of the
ratio $r_{s;i}/r_{e}$, on the basis of the data from \cite{mih09b}
about the mean potential energy of an electron in singly ionized
plasma. We remind that in \cite{mih09b} a quantity is determined
which is equal to $U_{p;c}/U_{D} = r_{D}/r_{c}$ (being just
designated differently), where $U_{D}= -e^2/r_{D}$ is the mean
potential energy determined within DH method \cite{deb23}. Keeping
in mind that $r_{D}= r_{e}/2^{1/2}$, and taking
$r_{c}/(\sqrt{2}\cdot r_{D}) = a_{c}$, we obtain $a_{c }=
r_{c}/r_{e}$, and then the expression (\ref{eq:pci}) for $r_{c}$
too. As, quantity $r_{D}/r_{c}$ is being determined among else in
\cite{mih09b}, just for the case of singly ionized plasma, so the
curve in Fig. 2 of this paper is obtained routinely from the curve
in Fig 4 in \cite{mih09b}, which represents quantity $r_{D}/r_{c}$
as a function of ratio $r_{s;i}/r_{e}$. The behaviour of $a_{c}$ in
a wide region of values of $r_{s;i}/r_{e}$ is presented in
Fig.\ref{fig::2}.

\subsection{Processes (\ref{eq:ph}) absorption coefficients.}

For the sake of further considerations it is necessary to have
energies $E_{n,l}$ of the bound states $|n,l>$ which are possible in
potential $U_{c}(r)$ and the corresponding radial wave functions
$R_{n,l}(r)$, as well as the radial wave functions $R_{E,l}(r)$
which correspond to the free states $|E,l>$. These quantities are
determined from Schrodinger equation
\begin{equation}
\label{eq:Shred} \frac{d^{2}R(r)}{dr^{2}} + \frac{2}{r}\frac{dR}{dr}
+ \left[\frac{2m}{\hbar^{2}}\cdot\left(W -U_{c}(r) \right) -
\frac{l(l+1)}{r^{2}}\right]= 0,
\end{equation}
where $W$ and $R(r)$ denote either $E_{n,l}$ and $R_{n,l}(r)$ or $E$
and $R_{E,l}(r)$. Here we will find the radial wave functions in the
form: $R_{n,l}(r) = P_{n,l}/r$ and $R_{E,l}(r) = P_{E,l}/r$, where
$P_{n,l}(r)$ and $P_{E,l}(r)$ can be expressed through the well known
analytical functions given in \cite{abr65}, namely:

 - in the region $0 < r < r_{c}$ all $P_{n,l}(r)$ and
$P_{E,l}(r)$ with $E < e^{2}/r_{c}$ are expressed through the
corresponding Witteker's functions, and $P_{E,l}(r)$ with $E >
e^{2}/r_{c}$ - through Coulomb-like ones which are regular at the
point $r = 0$;

 - in the region $r_{c} < r < \infty$ all $P_{n,l}(r)$ are
expressed through modified Bessel functions, and all $P_{E,l}(r)$ -
through spherical Bessel functions. Let us note that the bound-state
energies (in region $W < 0$) and the free-state phase shifts (in
region $W > 0$) are determined from the condition of continuity of
$R(r)$ and $dR(r)/dr$ at the point $r = r_{c}$.

Since the considered photo-ionization processes can be described in
the dipole approximation ($r_{c} << \lambda$), the corresponding
cross sections, for the non-perturbed bound states in potential
$U_{c}(r)$, are given by the expressions from \cite{sob79}, namely
\begin{equation}
\label{eq:sig} \sigma_{ph}(\lambda;n,l, E_{n,l}) = \frac{4 \pi ^2 e
^2 k_{ph}}{3 (2l + 1)} \displaystyle \sum _{l'=l \pm  1}
{l_{max}\left( \int\limits_{0}^{\infty} P_{n,l} r P_{E,l'} dr
\right) ^2},
\end{equation}
where  $k_{ph} = \varepsilon_{\lambda} / \hbar c$, $l_{max}\equiv
max(l,l')$, and $E = E_{n,l} + \varepsilon_{\lambda}$.

As it is well known, the spectral absorption coefficient $\kappa
_{ph} ^{(0)} (\lambda; N_{e}, T)$, characterizing the
photo-ionization processes (\ref{eq:ph}) in the case of
non-perturbed energy levels in the potential $U_{c}(r)$, is given by
the expression
\begin{equation}
  \label{eq:kappaph0}
  \kappa _{ph} ^{(0)} (\lambda; N_{e}, T) =
    \sum _{n,l}N_{n,l} \cdot \sigma_{ph}(\lambda;n,l,E_{n,l})\cdot
f_{st}(\lambda,T),
    \quad n \ge 2,
\end{equation}
where $\sigma_{ph}(\lambda;n,l,E_{n,l})$ is the partial
photo-ionization cross-section defined by Eq. \ref{eq:sig},
$N_{n,l}$ is the density of atoms $H^{*}(n,l)$, and factor $f_{st}$,
given by
\begin{equation}
  \label{eq:fst}
f_{st}(\lambda,T) = 1-exp(-2 \pi\hbar c/\lambda),
\end{equation}
describes the influence of the stimulated emission. One can see that
this expression is similar to the one for diluted hydrogen plasma
(see for example \cite{mih78}) and, in accordance with what was said
above, it cannot model the absorption coefficients of dense
non-ideal plasmas described in \cite{vit04}. In order to determine
the absorption coefficients $\kappa _{ph}(\lambda; N_{e}, T)$ which
can be applied to the mentioned modeling, we have to take into
account additional details of the atom-plasma interaction, beside
those which are already described by the shape of potential
(\ref{eq:U0}). It is known that within the usual way (i.e. as is
being done in the cases of weakly non-ideal plasmas) this additional
influence should be characterized by shifts and broadenings of the
bound-state energy levels of the considered atoms. However, this way
is generally applicable in any case, including the case of strongly
non-ideal plasma too.

This is confirmed by the approach of \cite{ada08}, main feature of
which is treatment of electrons in existing atoms $H^{*}(n,l)$ as a
Fermi gas of particles, which move in a self-consistent field
created by immobile ions and other electrons. One of the results of
this approach is a description of the relevant elementary event in
such a gas, i.e. annihilation of an electron localized at the $j$-th
proton in the chosen $\nu$-th state with a simultaneous creation of
an electron in some free state, caused by the absorption of a photon
with energy $\varepsilon_{\lambda}$. Namely, it was found that such
an event can be described in terms of the corresponding probability
density, which is practically equal to zero outside of a finite
interval of free-state energies. This result, despite the fact that
it has only qualitative significance, suggests that $\kappa
_{ph}(\lambda; N_{e}, T)$ can be obtained by a modification of
Eq.~(\ref{eq:kappaph0}) based on an adequately chosen probability
density $p_{n,l}(\epsilon)$ of the perturbed atom energy levels with
given $n$ and $l$, characterized by the corresponding shifts
$\Delta_{n,l}$ and broadenings $\delta_{n,l}$. It is assumed that
energies $\epsilon$ of the perturbed atomic states are dominantly
grouped around energy $\epsilon_{n,l}^{(max)}= E(n,l) +
\Delta_{n,l}$, inside the interval $(\epsilon_{n,l}^{(max)} -
\delta_{n,l}/2, \epsilon_{n,l}^{(max)} + \delta_{n,l}/2)$, similarly
to the known cases (Gaus, Lorentz, uniform etc.).

All this justifies a semi-empirical approach to the
considered problem, until a corresponding strict method is
developed. In this work any $(n,l)$-shell (with given $n$) of the
perturbed hydrogen atom is characterized by only two quantities,
namely an averaged shift $\Delta_{n}$ and broadening $\delta_{n}$,
which  are treated as empirical parameters. Consequently, here
it is considered that: $\Delta_{n,l} \equiv \Delta_{n}$,
$\delta_{n,l}\equiv \delta_{n}$ and $p_{n,l}(\epsilon)$ describe
the corresponding uniform distribution of the perturbed energy
levels, i.e.
\begin{equation}
\label{eq:pnl} p_{n,l}(\epsilon) = \left\{
\begin{array}{l}
\frac{1}{\delta_{n}},   \quad |\epsilon_{n,l}^{(max)} - \epsilon|
\le \frac{\delta_{n}}{2},
\\
\displaystyle 0 ,          \qquad |\epsilon_{n,l}^{(max)} -
\epsilon| > \frac{\delta_{n}}{2},
\end{array}
\right. \quad \epsilon_{n,l}^{(max)} = E_{n,l} + \Delta_{n}.
\end{equation}
Here we keep in mind that the uniform  distribution can
approximate many other distributions well (Gaussian, cupola etc.)

Let us note that although $\Delta_{n}$ is treated here as an
empirical parameter, it is possible to describe its qualitative
behaviour as a function of $N_{e}$. Namely, for well-known physical
reasons all shifts $\Delta_{n,l}$, and consequently $\Delta_{n}$,
have to change proportionally to the density of the perturbers, the
relative atom-perturber velocity and the characteristic perturbation
energy. Consequently, we will have it that
\begin{equation}
\label{eq:RelDelta} \Delta_{n} \cong Const. \cdot N_{e}\cdot
v_{ea}(T)\cdot e^{2}/l(N_{e},T),
\end{equation}
where $v_{ea}(T)$ and $l(N_{e},T)$ are the characteristic
electron-atom velocity and distance. On the basis of the results of
\cite{mih09b} in the considered cases ($N_{e}\sim 1\cdot 10^{19}
cm^{-3}$, $T \sim 2 \cdot 10^{4} K$) any relevant characteristic
length has to be close to the radius $r_{e}$, which is given by
Eq.~(\ref{eq:rei}). Since $v_{ea}(T)\sim (k_{B}T)^{1/2}$ and
$r_{e}\sim(k_{B}T/N_{e})^{1/2}$, from (\ref{eq:RelDelta}) the
relation follows
\begin{equation}
\label{eq:Deltan} \Delta_{n} \cong Const. \cdot N_{e}^{3/2},
\end{equation}
which is also in accordance with \cite{ada08}, and will be
particularly significant in further considerations.

Here, we will describe the perturbed atomic states in the first
order of the perturbation theory and, in accordance with what was said
above, we will have it that
\begin{equation}
\label{eq:kappaph} \kappa_{ph}(\lambda; N_{e}, T) =  \left[\sum
_{n,l}N_{n,l} \cdot \frac{\varepsilon_{\lambda}}{\delta_{n}} \int
\limits_{\epsilon^{-}}^ {\epsilon^{+}}
{\frac{\sigma_{ph}(\lambda^{(\epsilon)}; n, l,
E_{n,l})}{\varepsilon_{\lambda} + \epsilon} \cdot} d\epsilon\right]
\cdot f_{st}(\lambda, T),
\end{equation}
where $n \ge 2$, $\epsilon^{-}=
\epsilon_{n,l}^{(max)}-\delta_{n}/2$, $\epsilon^{+}=
\epsilon_{n,l}^{(max)}+\delta_{n}/2$, $\epsilon_{n,l}^{(max)} =
E_{n,l} + \Delta_{n}$, and $\sigma_{ph}(\lambda^{(\epsilon)};
n,l,E_{n,l})$ is given by Eq.~(\ref{eq:sig}) for
$\lambda^{(\epsilon)}= \lambda \cdot \varepsilon_{\lambda}/
(\varepsilon_{\lambda} + \epsilon)$, i.e. for the wavelength of a
photon with energy $(\varepsilon_{\lambda} + \epsilon)$.

\subsection{Processes (\ref{eq:inbs}), (\ref{eq:eH}) and (\ref{eq:HH}) absorption coefficients}

The spectral absorption coefficients characterizing the electron-ion
absorption process (\ref{eq:inbs}), electron-atom absorption
processes (\ref{eq:eH}) and ion-atom absorption processes
(\ref{eq:HH}) are determined here as in the previous papers
\cite{mih93,mih94,erm95}, dedicated to the same absorption processes
in some laboratory and astrophysical plasmas.

So, the spectral absorption coefficients $\kappa_{ei}(\lambda)$,
$\kappa_{ea}(\lambda)$ and $\kappa_{ia}(\lambda)$ are defined by
expressions
\begin{equation}
\label{eq:ei} \kappa_{ei}(\lambda) =  K_{ei}(\lambda, T)\cdot
N_{e}\cdot N_{i} \cdot f_{st}(\lambda, T) \cong K_{ei}(\lambda, T)
\cdot N_{e}^{2}\cdot f_{st}(\lambda, T),
\end{equation}
\begin{equation}
\label{eq:ea} \kappa_{ea}(\lambda) = \sigma_{phd}^{-} \cdot
S_{ea}^{-1} \cdot N_{e}\cdot N_{a} + K_{ea}(\lambda, T)\cdot
N_{e}\cdot N_{a}\cdot f_{st}(\lambda, T),
\end{equation}
\begin{equation}
\label{eq:S}  S_{ea}= \left (\frac{N_{e}\cdot N_{a}}{N_{neg.i}}
\right )_{eq} = 4 \left ( \frac{m k T}{2 \pi \hbar}\right
)^{\frac{3}{2}}\cdot e^{-\frac{E_{d}^{-}}{kT}},
\end{equation}
\begin{equation}
\label{eq:ia} \kappa_{ia}(\lambda) = K_{ia}(\lambda, T)\cdot
N_{i}\cdot N_{a}\cong K_{ia}(\lambda, T)\cdot N_{e}\cdot N_{a}\cdot f_{st}(\lambda, T),
\end{equation}
where $N_{a}$, $N_{i}$ and $N_{neg.i}$ are the densities of $H(1s)$
atoms, $H^{+}$ ions and negative $H^{-}(1s^{2})$ ions respectively,
"eq" denotes that the quantity $S_{ea}$ is determined under the
condition of thermodynamic equilibrium of the considered system,
$E_{d}^{-}$ is the energy of $H^{-}(1s^{2})$ ion dissociation, and
factor $f_{st}(\lambda, T)$ is given by expression (\ref{eq:fst}).

The spectral coefficient $K_{ei}(\lambda, T)$, given in $[cm^{5}]$,
is calculated by means of the corresponding expressions from
\cite{sob79}, with the Gaunt factor determined in \cite{joh72}. The
spectral coefficient $K_{ea}(\lambda, T)$, given also in $[cm^{5}]$,
and the cross-section $\sigma_{phd}^{-}$ for photo-dissociation of
ion $H^{-}(1s^{2})$ are determined on the basis of the expressions
from \cite{sti70} and \cite{wis79}. Finally, the spectral
coefficient $K_{ia}(\lambda, T)$ is determined in the quasi-static
approximation, which is described in detail in \cite{erm95}, and is
taken in the form
\begin{equation}
\label{eq:Kia} K_{ia}(\lambda,T) = 0.62\times10^{-42}
\frac{C(R_{\lambda})(R_{\lambda}/a_{0})^{4}}{1-a_{0}/R_{\lambda}}
\cdot exp \left (- \frac{U_{1}(R_{\lambda})}{kT} \right ),
\end{equation}
where $U_{1}(R)$ is the energy of the electronic ground state $1
\Sigma_{g}^{+} $ of molecular ion $H_{2}^{+}$, as a function of
internuclear distance $R$, $R_{\lambda}$ - the resonance
internuclear distance for given $\lambda$, $a_{0}$  - the atomic
unit of length, and $C(R_{\lambda})$ - a dimensionless coefficient
which is close to one. Parameters $R_{\lambda}$,
$U_{1}(R_{\lambda})$ and $C(R_{\lambda})$, as functions of
$\lambda$, are tabulated in \cite{mih93,erm95}. The relative
contributions of the partial channels, i.e. $H(1s)$-$H^{+}$
absorption charge exchange and molecular ion
$H_{2}^{+}(1\Sigma^{+}_{g})$ photo-dissociation, are obtained by
multiplication of $K_{ia}(\lambda, T)$ by factors $X(z)$ and
$[1-X(z)]$ respectively. In the considered region of $\lambda$
factor $X(z) = \Gamma(3/2;z)/\Gamma(3/2)$, where $z =
-U_{1}(R_{\lambda})/kT$.

\section{Results and discussion \label{sec:rd}}

\subsection{The characteristics of the cut-off Coulomb potential}

In this paper the approximation of cut-off Coulomb potential
(\ref{eq:U0}) is applied to modeling the spectral absorption
coefficient obtained in \cite{vit04} in two experiments with
hydrogen plasmas, which are treated as a short and a long pulse
respectively. In the first case (short pulse) plasma with $N_e= 1.5
\cdot 10^{19} cm^{-3}$ and $T = 2.3\cdot 10^{4} K$ was studied,
while in the second case (long pulse) - it was plasma with $N_{e}=
6.5 \cdot 10^{18} cm^{-3}$ and $T =1.8\cdot 10^{4} K$. In the
experiments described in \cite{vit04} plasmas with electron
densities up to $\approx 10^{19} cm^{-3}$ were created by pulse
discharge in quartz capillary. Diagnostics of the plasma was carried
out on the basis of optical measurements (at $\lambda = 632.8 nm$),
taking into account radial inhomogeneity of the plasma column. The
temperature profile is defined from independent measurements of
brightness and transparency at different distances from the center
of the capillary. A detailed study is performed just for the two
above mentioned examples.

\begin{figure}[h!]
\centering
      \includegraphics[width=\columnwidth,
height=0.75\columnwidth]{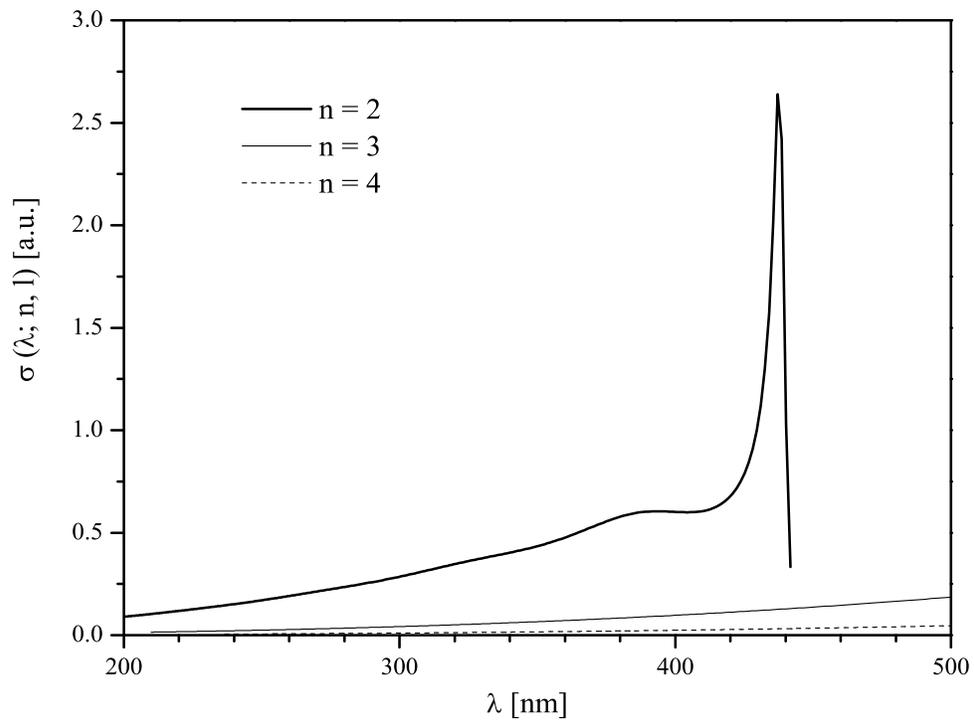} \caption{The
photo-ionization cross-section for the bound state with $l=0$ and
$n=2$, $3$ and $4$ in the potential $U_{c}(r)$ for short pulse
($r_{c}=44.964a.u.$).} \label{fig::3}
\end{figure}

\begin{figure}[h!]
\centering
      \includegraphics[width=\columnwidth,
height=0.75\columnwidth]{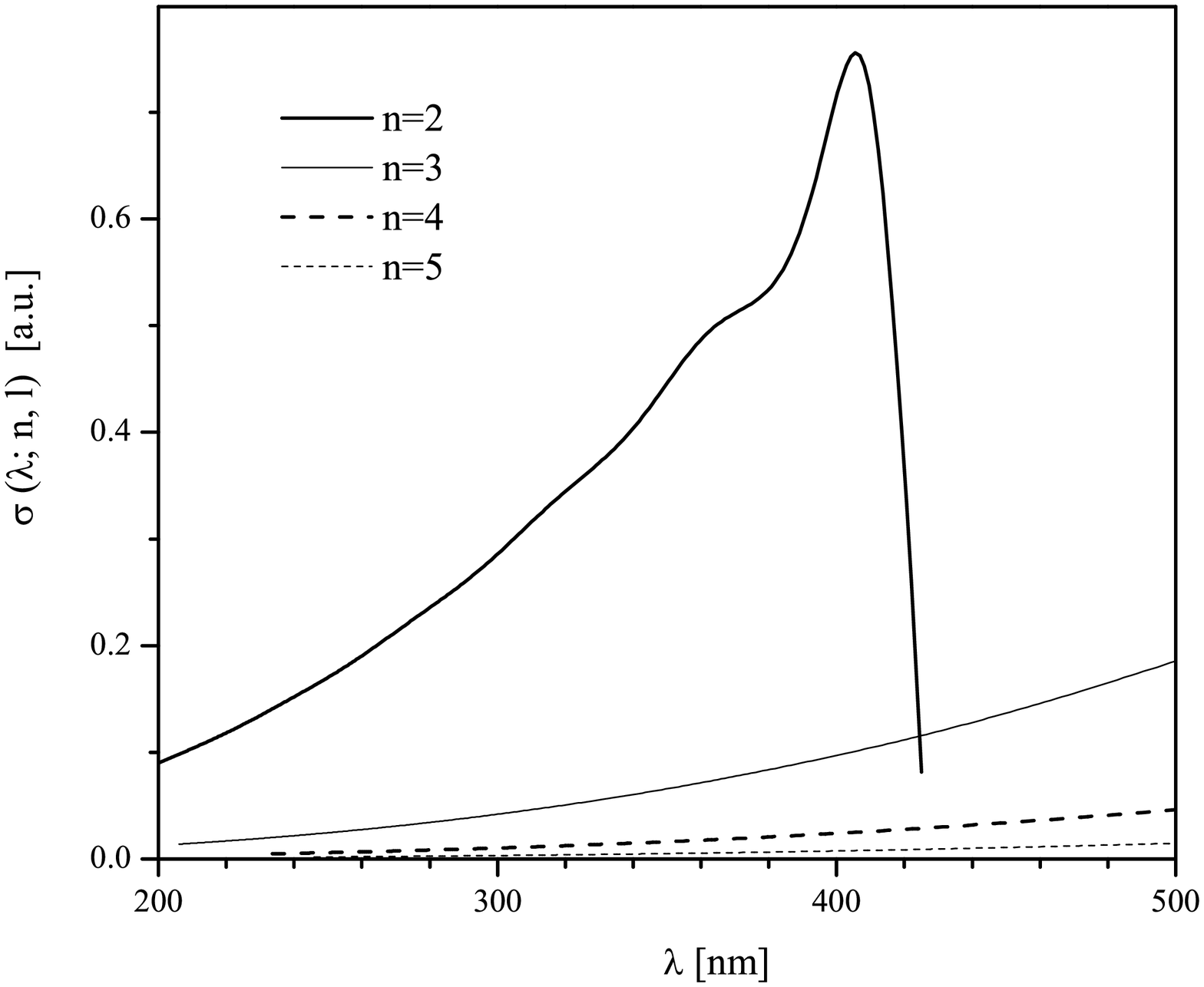} \caption{The photo-ionization
cross-section for the bound state with $l=0$ and $n=2$, $3$, $4$ and
$5$ in the potential $U_{c}(r)$ for the long pulse ($r_{c}=55.052
a.u.$).} \label{fig::4}
\end{figure}

On the basis of Fig.\ref{fig::2} and Eq.~(\ref{eq:pci}) it was found
that the cut-off radius $r_{c}$ is equal to : $44.964$ a.u. for the
short pulse, and $55.052$ a.u. for the long one. For these values of
$r_{c}$ the solutions of Eq.~(\ref{eq:Shred}) correspond to the
energies of the realized bound state, which are presented in Tab's 1
and 2 respectively. The corresponding partial photo-ionization
cross-sections $\sigma(\lambda;n,l,\epsilon_{n,l})$ are obtained by
means of Eq.~(\ref{eq:sig}) for $n$, $l$ and $\epsilon_{n,l}$ given
in  Tab's 1 and 2. The behavior of these cross-sections is
illustrated in Fig.\ref{fig::3} and Fig.\ref{fig::4} by the examples
of photo-ionization cross-sections of all realized states with l=0.
These figures show qualitative similarity of behavior of the
cross-sections in the cases $r_{c}= 44.964 a.u.$ and  $ r_{c}=
55.052 a.u.$ and domination of the cross-sections with n=2. One can
see a significant difference between the maximal values of the
cross-sections with n=2 (about 2.70 a.u. and 0.75 a.u.) which
correspond to these cases. This fact reflects the tendency of a
significant decrease of the maximal values of the cross-sections for
n=2 with an increase of the cut-off radius $r_{c}$.

Let us remind that Tab's 1 and 2 characterize the bound states of an
electron in the potential $U_{c}(r)$ with the values of cut-off
radius $r_{c}$ given above. The energies of the corresponding ground
states approach the value of $-I_{H}$, where $ I_{H}=13.598 eV$ is
the tabulated value of the isolated hydrogen-atom ionization
potential (see for an example NIST Atomic Spectra Database), only
when $r_{c}\rightarrow \infty$, i.e. when $N_e\rightarrow 0$ or $T
\rightarrow \infty$. Also, the energies of the ground states (for
the electron densities and the temperatures observed) would be close
to ($-I_{H}$) in the case where instead of $U_{c}(r)$ the potential
$U(r; r_{c})$ would be used (see Fig.~1).

For each of the considered cut-off radii $r_{c}$ the existence is
assumed here of a Boltzmann's distribution of the populations
$N_{n,l}$ of the bound states, given in Tab 1 or 2, which exist in
Eq. (\ref{eq:kappaph}). Such a distribution is determined by the
corresponding values of the total density of neutral hydrogen atoms
$N_{a}$ and the temperature $T$. In accordance with \cite{vit04}
here it is taken that: $N_{a}=1.9\cdot10^{19} cm^{-3}$ and $T=22980
K$ for $r_{c}= 44.964 a.u.$, and $N_{a}=3.4\cdot10^{19} cm^{-3}$ and
$T=17960 K$ for $ r_{c}= 55.052 a.u.$.  As one of the consequences,
we have it that the total populations of groups of the states with
same $n$ are equal to: $4.4\cdot10^{17}$, $3.82 \cdot10^{17}$ and
$4.86 \cdot10^{17} cm^{-3}$ for $n=2$, $3$ and $4$ in the first
case, and $1.87 \cdot10^{17}$,  $1.24 \cdot10^{17}$, $1.44
\cdot10^{17}$ and $1.85 \cdot10^{17} cm^{-3}$ for $n=2$, $3$ , $4$
and $5$ in the second case.

\subsection{The absorption coefficient: the results of the calculations}

In order to compare the obtained theoretical results with the
experimental data from \cite{vit04}, we have to take into account
all the absorption processes which cannot be neglected in the
considered hydrogen plasmas, i.e. the processes described by
Eqs.~(\ref{eq:ph}), (\ref{eq:inbs}), (\ref{eq:eH}) and
(\ref{eq:HH}). Therefore, when comparing our theoretical results
with the experimental data from \cite{vit04} we have to use the
corresponding total spectral absorption coefficient
$\kappa_{tot}(\lambda)$, namely
\begin{equation}
\label{eq:tot} \kappa_{tot}(\lambda) = \kappa_{ph}(\lambda) +
\kappa_{add}(\lambda),
\end{equation}
where $\kappa_{ph}(\lambda)\equiv \kappa_{ph}(\lambda;N_{e},T)$ is
given by Eq.~(\ref{eq:kappaph}), and the member
$\kappa_{add}(\lambda) \equiv \kappa_{add}(\lambda;N_{e},N_{a},T)$
characterizes the contribution of absorption processes
(\ref{eq:inbs}), (\ref{eq:eH}) and (\ref{eq:HH}). Consequently, we
have it that
\begin{equation}
\label{eq:add} \kappa_{add}(\lambda) = \kappa_{ei}(\lambda) +
\kappa_{ea}(\lambda) + \kappa_{ia}(\lambda),
\end{equation}
where $\kappa_{ei}(\lambda)\equiv\kappa_{ei}(\lambda;N_{e},T)$,
$\kappa_{ea}(\lambda)\equiv\kappa_{ea}(\lambda; N_{e},N_{a},T)$ and
$\kappa_{ia}(\lambda)\equiv\kappa_{ia}(\lambda;N_{e},N_{a},T)$ are
the partial spectral absorption coefficients, which are given above
by Eqs.~(\ref{eq:ei}), (\ref{eq:ea}), (\ref{eq:S}), (\ref{eq:ia})
and (\ref{eq:Kia}).

In accordance with the aims of this work the calculations of the
total absorption coefficients  $\kappa_{tot}(\lambda)$ have been
performed for both cases (short and long pulse) in a wide region of
values of shifts ($\Delta_{n}$) and broadenings ($\delta_{n}$) of
atomic levels with $n \ge 2$. The calculations of
$\kappa_{tot}(\lambda)$ cover the wavelength region $300nm \le
\lambda \le 500nm$. However, let us emphasize the fact that the
values of the experimental absorption coefficient
$\kappa_{exp}(\lambda)$ characterize not only the bound-free
(photo-ionization) processes (\ref{eq:ph}) and the said additional
absorption processes (\ref{eq:inbs}) - (\ref{eq:HH}), but also the
bound-bound (photo-excitation) processes, which are not considered
in this work. Consequently, for the purpose of this work the region
$\lambda \lesssim 450nm$ in the case of short pulse, and $\lambda
\lesssim 425nm$ in the case of long pulse, (see Tabs.~1 and 2) has
the real significance, where the considered photo-ionization
processes dominate in comparison with photo-excitation ones. The
results of calculations are shown in Figs.~5-10 together with the
corresponding experimental values $\kappa_{exp}(\lambda)$ of the
spectral absorption coefficient from \cite{vit04}.

Figures 5, 6, 7 and 8 illustrate the results of the calculations
of $\kappa_{tot}(\lambda)$ in the case when $\Delta_{n} = const.$,
while Figs.~9 and 10 present the calculations of
$\kappa_{tot}(\lambda)$ in the case when $\Delta_{n}$ decreases
(relative to $\Delta_{2}$) with increasing $n$.

The bottom and the top curves of $\kappa_{tot}(\lambda)$ in Figs.~5
and 6 illustrate strong influence of $\Delta_{n}$ on the calculated
total absorption coefficient: $\Delta_{n}=\delta_{n}=0.30eV$ and
$\Delta_{n}=\delta_{n}=0.60eV$ for the short pulse;
$\Delta_{n}=\delta_{n}=0.05eV$ and $\Delta_{n}=\delta_{n}=0.20eV$
for the long pulse. The groups of three curves, which lie between
the corresponding bottom and top curves, demonstrate relatively
small influence of $\delta_{n}$ on the calculated values of
$\kappa_{tot}(\lambda)$: $\Delta_{n}=0.45eV$ and
$\delta_{n}=0.40eV$, $\delta_{n}=0.45eV$ and $\delta_{n}=0.50eV$ for
the short pulse; $\Delta_{n}=0.125eV$ and $\delta_{n}=0.120eV$,
$\delta_{n}=0.125eV$ and $\delta_{n}=0.130eV$ for the long pulse.

Also, the dashed curves on Figs.~5 and 6 demonstrate the behaviour
of the spectral absorption coefficient $\kappa_{add}(\lambda)$,
defined by Eqs. (\ref{eq:add}) and (\ref{eq:ei}) - (\ref{eq:Kia}),
in the case of short and long pulses respectively. One can see that
the total contribution of electron-ion, electron-atom and ion-atom
absorption processes, described by Eqs.~(\ref{eq:inbs}),
(\ref{eq:eH}) and (\ref{eq:HH}) indeed cannot be neglected in the
considered cases.

Figures 7 and 8 show the curves of $\kappa_{tot}(\lambda)$
calculated with the values of $\Delta_{n}$ and $\delta_{n}$ which
are treated as the optimal ones: $\Delta_{n}= 0.455 eV$ and
$\delta_{n}= 0.625 eV$ for the short pulse, and $\Delta_{n}=0.13 eV$
and $\delta_{n}= 0.11 eV$ for the long one. In order to estimate the
possible error due to such a choice of shifts and broadenings, the
results of calculations are shown in Figs.~9 and 10 in the case when
$\Delta_{n}$ decreases with the increase of $n$, proportionally to
the ionization energies of the corresponding atomic levels. Let us
note the fact that calculated curves presented in these figures
correspond to the optimal values of $\Delta_{n=2}$ and
$\delta_{n=2}$. One can see that the calculated curves in Figs.~9
and 10 are very close to the calculated curves in Figs.~7 and 8,
respectively. This fact is reflected in the values of $\Delta_{n=2}$
and $\delta_{n=2}$ which correspond to the curves in Figs.~9 and 10:
$\Delta_{n=2}=0.49 eV$ and $\delta_{n=2}=0.65 eV$ for the short
pulse, and $\Delta_{n=2}=0.14 eV$ and $\delta_{n=2}=0.12 eV$ for the
long one.

Beside the curves $\kappa_{tot}(\lambda)$ and
$\kappa_{exp}(\lambda)$, the curves $\kappa_{p.th.}(\lambda)$ are
also presented in Figs.~7 and 8, which were obtained in \cite{vit04}
for the short and long pulse on the basis of the perturbation theory
used in \cite{gav01,vit04}. For the sake of correct interpretation
of the presented data, let us note the fact that the values of
$\kappa_{p.th.}(\lambda)$, similarly to $\kappa_{exp}(\lambda)$,
characterize not only photo-ionization, but also the bound-bound
(photo-excitation) processes, which are not considered in this work.
One can see that in the case of strongly non-ideal plasmas (short
pulse, $N_{e}=1.5\cdot 10^{19} cm^{-3}$) the difference between
$\kappa_{exp}(\lambda)$ and $\kappa_{p.th.}(\lambda)$ is so large
that it justifies any effort towards development of an alternative
method of calculation of the strongly non-ideal plasma absorption
coefficient. However, in the case of long pulse ($N_{e}=0.65\cdot
10^{19} cm^{-3}$) the considered plasma is located by its parameters
in the lower part of the region of strong non-ideality, where the
perturbation theory should give much better results. This fact is
reflected in a significant reduction of the difference between
$\kappa_{exp}(\lambda)$ and $\kappa_{p.th.}(\lambda)$.

Therefore it is important to check whether relation
Eq.~(\ref{eq:Deltan}) is valid also for $N_{e}$ close to $0.65\cdot
10^{19} cm^{-3}$. Since in the case of constant shifts
$\Delta_{n}=0.455 eV$ and $0.130 eV$ for short and long pulses
respectively, validity of Eq.~(\ref{eq:Deltan}) means that
$0.455/0.130=(1.5/0.65)^{3/2}$, which is satisfied with an accuracy
better than $1\%$. In the case of variable shift we have it that
$\Delta_{n=2}=0.49 eV$ and $0.12 eV$ for the short and long pulse
respectively, and validity of Eq.~(\ref{eq:Deltan}) means now that
$0.49/0.14=(1.5/0.65)^{3/2}$, which is satisfied with the same
accuracy.

This fact offers a possibility to determine $\Delta_{n}$ or
$\Delta_{n=2}$ for any $N_{e}$ and $T$ from intervals $0.65 \cdot
10^{19} cm^{-3} < N_{e} < 1.5\cdot 10^{19} cm^{-3}$ and $1.8 \cdot
10^{4} K \le T \le 2.3 \cdot 10^{4} K$, and probably in some
significantly wider regions, at least for $5\cdot 10^{18} cm^{-3}
\lesssim N_e \lesssim 5 \cdot 10^{19}cm^{-3}$ and $1.6 \cdot 10^{4}
K \lesssim T \lesssim 2.5 \cdot 10^{4} K$. Then, since it has been
established that the influence of $\delta_{n}$ is significantly
weaker than the influence of $\Delta_{n}$, we can determine
$\delta_{n}$ for any $N_{e}$ from those intervals, taking
$\delta_{n} = \Delta_{n}$, as in the examples illustrated by the
dashed curves on Figs.~\ref{fig::7} and \ref{fig::8}.

\centerline{***}

On the grounds of all that has been said one can conclude that the
presented method can already be used for calculations of the
spectral absorption coefficients of dense hydrogen plasmas in the
regions $N_{e}\sim 10^{19} cm^{-3}$ and $T_{e}\approx 2 \cdot
10^{4}K$, as long as electron-$H^{+}$ and electron-$H(1s)$ inverse
bremsstrahlung, negative ion $H^{-}(1s^{2})$ photo-detachment,
$H(1s)$-$H^{+}$ absorption charge exchange and molecular ion
$H_{2}^{+}(1\Sigma^{+}_{g})$ photo-dissociation can be described as
in this paper. Let us note the fact that with some modifications
related to the atom photo-ionization processes, which should enable
description of the influence of the atom core, the presented method
can be applied to other kinds of laboratory dense hydrogen-like
plasmas. Most of all, we mean alkali metals, helium and other
rare-gas plasmas. Also, the obtained results can be of interest for
some astrophysical plasmas, namely the plasma of the inner layers of
solar atmosphere, as well as the plasmas of partially ionized layers
of some other stellar atmospheres (for example some DA and DB white
dwarfs).

Further development of the method described requires first of all an
improvement of the procedure used, in order to replace the
semi-empirical parameters with ones determined within the procedure
itself. Another step would be including into consideration atom
bound-bound (photo-excitation) processes which have been omitted
here, and extending the region of the method's applicability to the
long wavelengths.

\section{Acknowledgments}

The authors would like to express their gratitude to Prof. V.~M.~
Adamyan and Prof. Lj.~M.~Ignjatovi{\' c} for their permanent
attention, support and useful discussions. The authors are thankful
to the University P. et M. Curie of Paris (France) for financial
support, as well as to the Ministry of Science of the Republic of
Serbia for support within the Project 176002 "Influence of
collisional processes on astrophysical plasma line shapes".

%

\begin{table}[h]
\caption{The energies of the bound states in the potential
$U_{c}(r)$ in the case of short pulse (cut-off radius $r_{c}=44.964$
a.u.): the principal and orbital  quantum numbers $n$ and $l$, and
the corresponding energies $E_{n,l}$ in [$cm^{-1}$].}
\begin{center}
\begin{tabular}
{l c c c c } \hline
   & \multicolumn{4}{c}{l}\\
   \cline{2-5}
  n & 0& 1& 2& 3\\
\hline
  1 & -104856.11387 & & &\\
  2 & -22553.19871 & -22553.19871 & &\\
  3 & -7311.91633 & -7311.91633 & -7311.91633 &\\
  4 & -1979.09916 & -1978.63943 & -1978.01838  & -1977.59897 \\
\hline
\end{tabular}
\end{center}
\label{tab:dva}
\end{table}

\begin{table}[h]
\caption{The energies of the bound states in the potential
$U_{c}(r)$ in the case of long pulse (cut-off radius $r_{c}=55.052$
a.u.): the principal and orbital  quantum numbers $n$ and $l$, and
the corresponding energies $E_{n,l}$ in [$cm^{-1}$].}
\begin{center}
\begin{tabular}
{l c c c c c } \hline
   & \multicolumn{5}{c}{l}\\
   \cline{2-6}
  n & 0& 1& 2& 3& 4\\
  \hline
  1 & -105750.58278 & & & &\\
 2 & -23447.66762 & -23447.66762 & & &\\
 3 & -8206.36911 & -8206.36911 & -8206.36911 & &\\
 4 & -2871.96303 & -2871.94689 & -2871.93883 & -2871.93076 &\\
 5 & -428.40705 & -423.27495 & -415.40217 & -408.2198 & -404.06201 \\
\hline
\end{tabular}
\end{center}
\label{tab:jedan}
\end{table}

\begin{figure}[h!]
\centering
      \includegraphics[width=\columnwidth,
height=0.75\columnwidth]{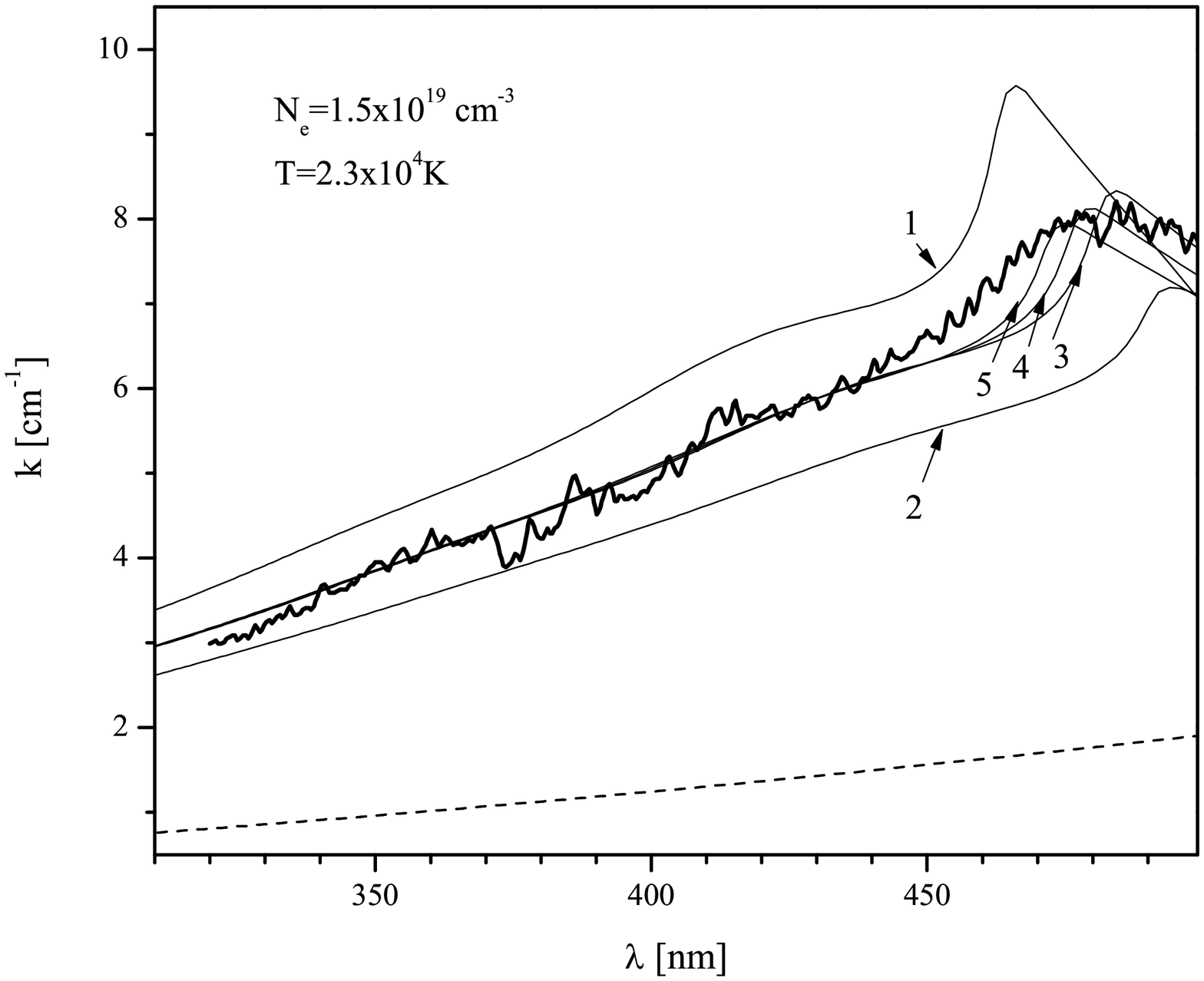} \caption{The influence
of the shift and broadening to the calculated total absorption
coefficient. Short pulse: 1 - absorption coefficient
$\kappa_{tot}(\lambda)$ calculated in the approximation of the
constant shift with $\Delta_{n}=\delta_{n}=0.3eV$; 2- the same, but
with $\Delta_{n}=\delta_{n}=0.6eV$; 3, 4 and 5 - the same, but with
$\Delta_{n}=0.45eV$ and $\delta_{n}=0.4eV$, $0.45eV$ and $0.5eV$
respectively. Dashed curve - absorption coefficient
$\kappa_{add}(\lambda)$ which characterizes the total contribution
of absorption processes (\ref{eq:inbs}) - (\ref{eq:HH}). Bold curve
- the experimental absorption coefficient from \cite{vit04}.}
\label{fig::5}
\end{figure}

\begin{figure}[h!]
\centering
      \includegraphics[width=\columnwidth,
height=0.75\columnwidth]{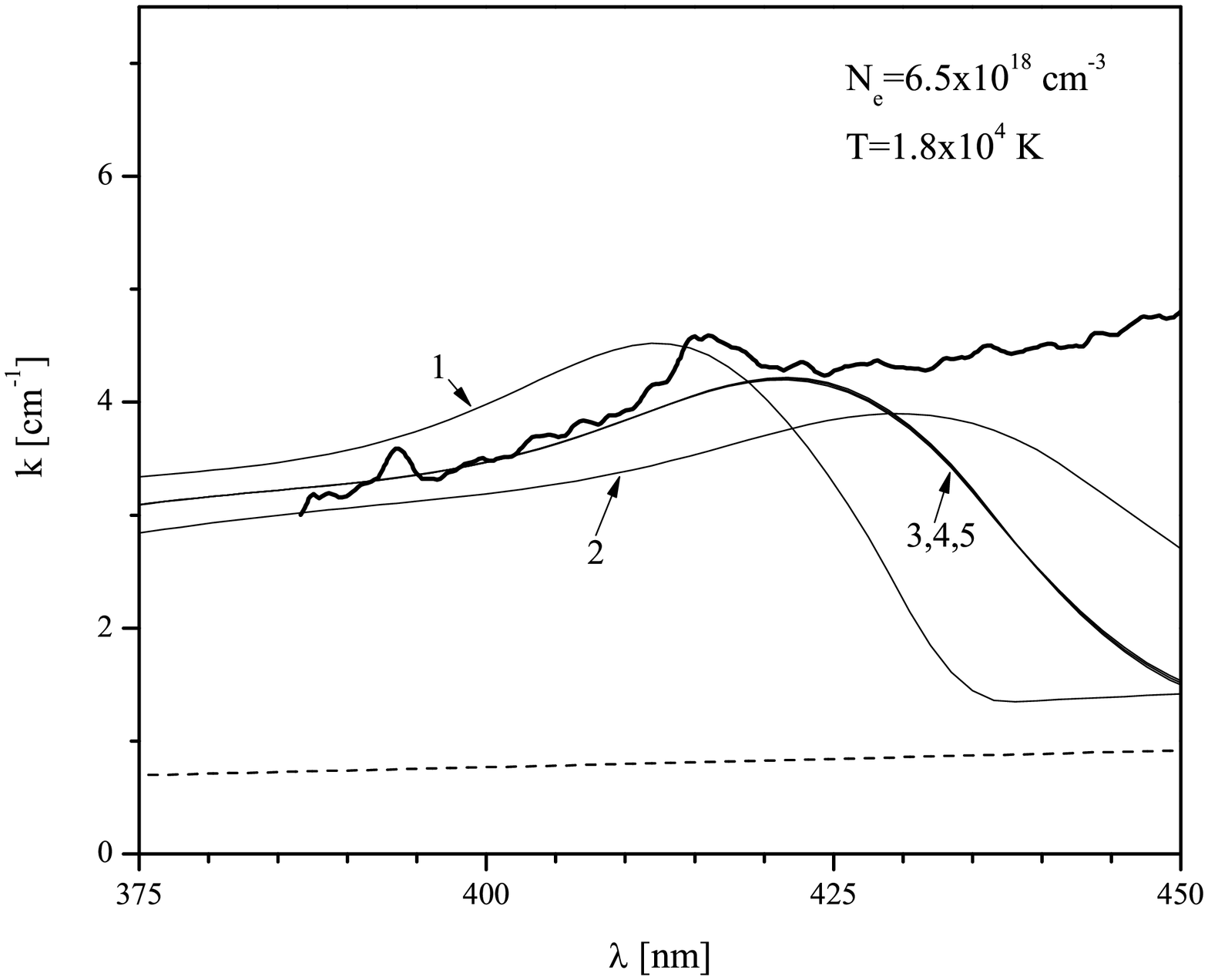} \caption{The influence
of the shift and broadening to the calculated total absorption
coefficient. Long pulse: 1 - absorption coefficient
$\kappa_{tot}(\lambda)$ calculated in the approximation of the
constant shift with $\Delta_{n}=\delta_{n}=0.05eV$; 2 - the same,
but with $\Delta_{n}=\delta_{n}=0.2eV$; 3, 4 and 5 - the same, but
with $\Delta_{n}=0.125eV$ and $\delta_{n}=0.12eV$, $0.125eV$ and
$0.13eV$ respectively. Dashed curve - absorption coefficient
$\kappa_{add}(\lambda)$ which characterizes the total contribution
of absorption processes (\ref{eq:inbs}) - (\ref{eq:HH}). Bold curve
- the experimental absorption coefficient from \cite{vit04}.}
\label{fig::6}
\end{figure}

\begin{figure}[h!]
\centering
      \includegraphics[width=\columnwidth,
height=0.75\columnwidth]{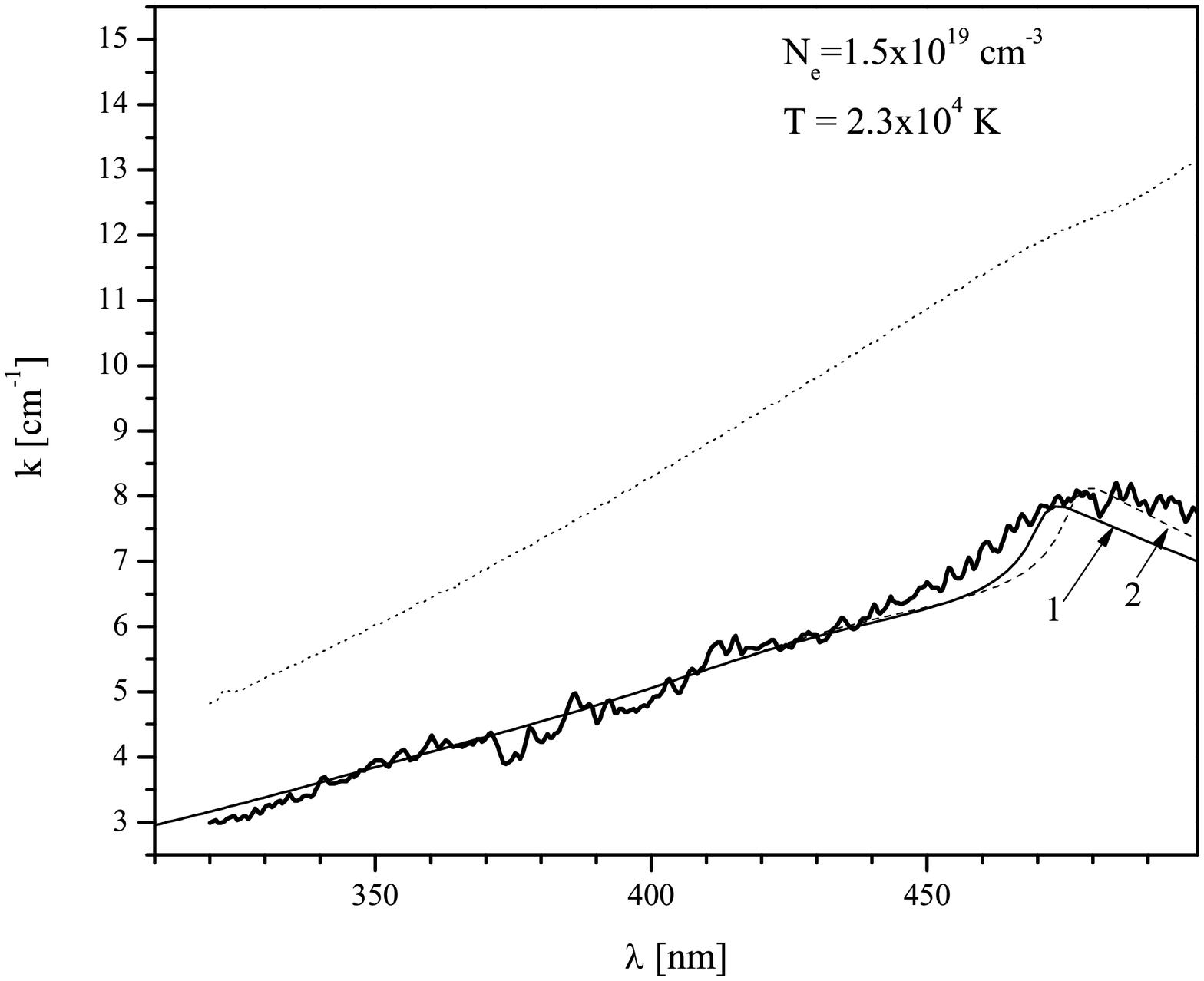} \caption{Short pulse:
1 - absorption coefficient $\kappa_{tot}(\lambda)$ calculated in the
approximation of the constant shift with the optimal values of
$\Delta_{n}$ and $\delta_{n}$, namely $\Delta_{n}=0.455eV$ and
$\delta_{n}=0.525eV$; 2 - absorption coefficient
$\kappa_{tot}(\lambda)$ calculated in the same approximation with
$\Delta_{n}= \delta_{n} =0.455eV$.   Bold and doted curve - the
absorption coefficients obtained in \cite{vit04} experimentally and
by means of the perturbation theory.} \label{fig::7}
\end{figure}

\begin{figure}[h!]
\centering
      \includegraphics[width=\columnwidth,
height=0.75\columnwidth]{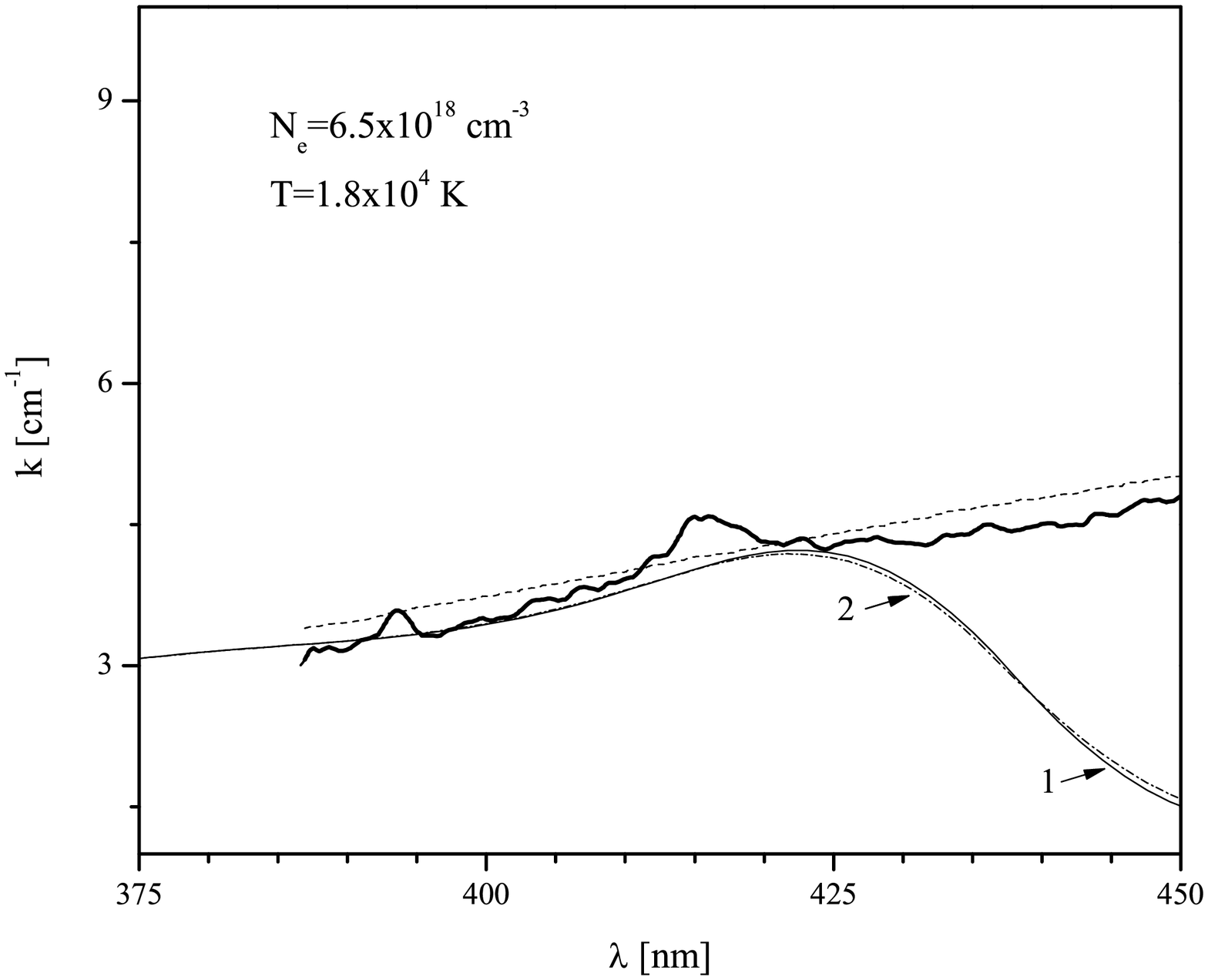} \caption{Long pulse: 1
- absorption coefficient $\kappa_{tot}(\lambda)$ calculated in the
approximation of the constant shift with the optimal values of
$\Delta_{n}$ and $\delta_{n}$, namely $\Delta_{n}=0.13eV$ and
$\delta_{n}=0.11eV$; 2 - absorption coefficient
$\kappa_{tot}(\lambda)$ calculated in the same approximation with
$\Delta_{n}=\delta_{n}=0.13eV$. Bold and dashed line - the
absorption coefficients obtained in \cite{vit04} experimentally and
by means of the perturbation theory.} \label{fig::8}
\end{figure}

\begin{figure}[h!]
\centering
      \includegraphics[width=\columnwidth,
height=0.75\columnwidth]{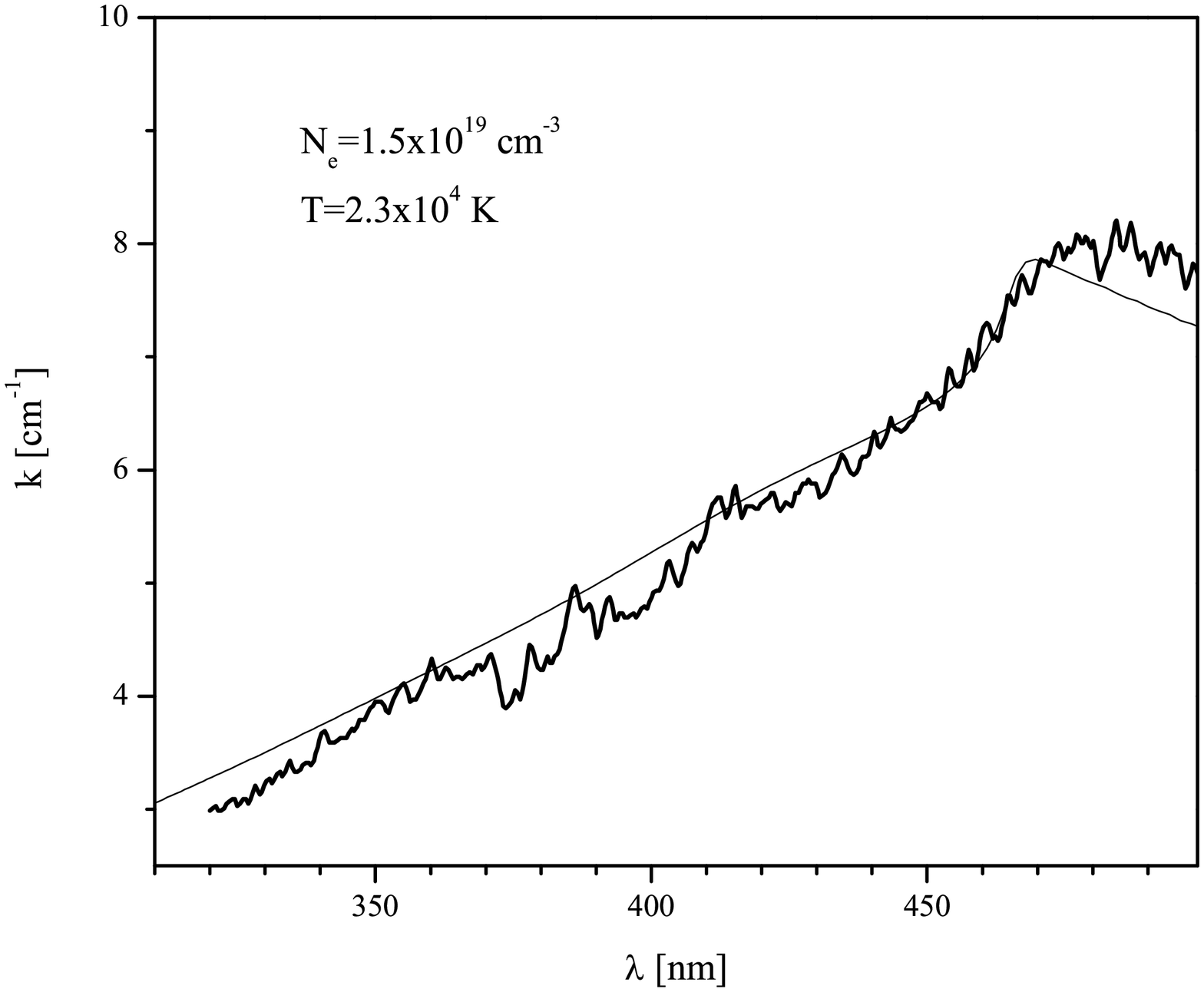} \caption{Short pulse:
thin line - absorption coefficient $\kappa_{tot}(\lambda)$
calculated in the approximation of the variable shift with the
optimal values of $\Delta_{n=2}$ and $\delta_{n=2}$, namely
$\Delta_{n=2}=0.49eV$ and $\delta_{n=2}=0.55eV$. Bold line - the
experimental absorption coefficient from \cite{vit04}.}
\label{fig::9}
\end{figure}

\begin{figure}[h!]
\centering
      \includegraphics[width=\columnwidth,
height=0.75\columnwidth]{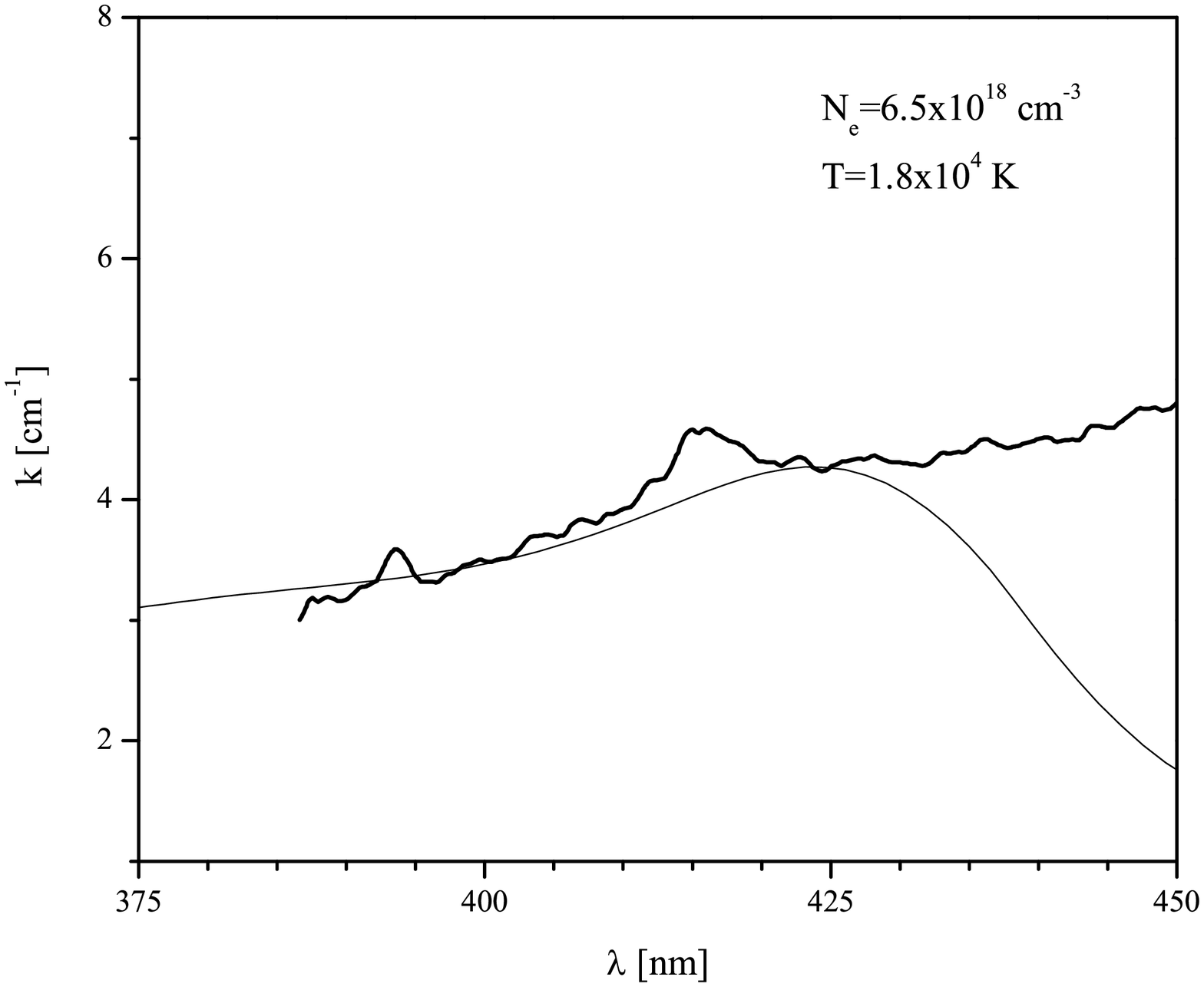} \caption{Long pulse:
thin line - absorption coefficient $\kappa_{tot}(\lambda)$
calculated in the approximation of the variable shift with the
optimal values of $\Delta_{n=2}$ and $\delta_{n=2}$, namely
$\Delta_{n=2}=0.14eV$ and $\delta_{n=2}=0.12eV$. Bold line - the
experimental absorption coefficient from \cite{vit04}.}
\label{fig::10}
\end{figure}

\end{document}